\documentclass[12pt]{iopart}
\usepackage{graphicx}
\begin{document}

\title[Evaluation of Tranche in Securitization]
{Evaluation of Tranche in Securitization and  
Long-range Ising Model}

\author{K Kitsukawa \dag \footnote[1]{kj198276@sfc.keio.ac.jp}
, S Mori \ddag 
\footnote[2]{mori@sci.kitasato-u.ac.jp} and 
M Hisakado \P \footnote[3]{masato\_hisakado@standardpoors.com}   
}

\address{\dag\  Graduate School of Media and Governance, Keio University, 
Endo 5322, Fujisawa, Kanagawa 252-8520, Japan}

\address{\ddag\ Department of Physics, School of Science,
Kitasato University, Kitasato 1-15-1 , Sagamihara, Kanagawa 228-8555, Japan}

\address{\P\ 
Standard \& Poor's, Marunouchi 1-6-5, Chiyoda-ku, Tokyo 100-0005, Japan}

\vspace*{6cm}

\begin{abstract}
This econophysics work  studies the long-range Ising model
 of a finite system with $N$ spins and 
the exchange interaction $\frac{J}{N}$ and 
the external field $H$ as a model
for homogeneous credit portfolio of assets with  
default probability $P_{d}$ and 
default correlation $\rho_{d}$. 
Based on the discussion on  the $(J,H)$ phase diagram, 
 we develop a perturbative calculation method for the model and  
obtain explicit expressions for $P_{d},\rho_{d}$ and the normalization 
factor $Z$ in terms of the 
model parameters $N$ and $J,H$.   
The effect of the default correlation $\rho_{d}$
 on the probabilities $P(N_{d},\rho_{d})$ for $N_{d}$ defaults  and on the
cumulative distribution function $D(i,\rho_{d})$ are discussed.
The latter means the average loss rate of the``tranche'' 
(layered structure ) of the
securities (e.g. CDO), which are synthesized from a pool of many 
assets. We show that the expected loss rate of the 
subordinated tranche decreases with $\rho_{d}$ and  that of the senior
tranche increases linearly, which are important in their 
pricing and ratings. 
\end{abstract}


\submitto{Physica A}

\maketitle

\section{Introduction}

The statistical properties of the models for credit risks 
have been widely discussed in the past ten years from the standpoint of
financial
engineering \cite{Fabozzi,Schonbucher} and econophysics
\cite{Bouchaud,Stanley2}. 
In the context of econophysics, the mechanism of systemic failure in
banking has been studied \cite{Aleksiejuk,Iori2}. Power law behavior of the
distributions of 
avalanches and several scaling laws in the context of percolation theory
were found. 
On the other hand, in
financial engineering, the
evaluation of the effect of the 
correlation $\rho$ between the rates of return of assets  or
between the default of assets is a
 hot topic and is widely discussed from theoretical 
and empirical
 viewpoints. Empirically, historically realized values of correlations  
and their implied values, which are estimated based on the 
market value 
 of credit derivatives,  
 are compared and their discrepancies, called correlation risk premium,
 attract investors' interests  from  the viewpoint of portfolio
 management \cite{Calamaro}.  
Theoretically, many statistical models are proposed for modeling  
credit risk of the pool of many assets 
\cite{Schonbucher,Merton,Cifuentes,Witt,Molins,Martin,
Finger,Li,Duffie,Duffie2}. 
There are two categories in the models. The models in the 
first category use two state discrete variables which describe whether the 
asset is defaulted or not \cite{Cifuentes,Witt,Aleksiejuk,Molins}.  
In the financial literature a
 two-valued variable $x_{i} (i=1,\cdots,N)$ 
takes values $0$ and $1$ depending on whether 
the $i$-th asset described by $x_{i}$ is not defaulted  or defaulted.  
The default probability $P_{d}$ is defined by the average number of 
defaulted assets $N_{d}$ per an asset as $P_{d}=\frac{<N_{d}>}{N}$.
Here $<\hspace*{0.5cm}>$ means the expectation value.
Ising Spin variable
$S=\pm 1$ is also used and  it is related to $x$ as $S=1-2 x$.
Moody's Binomial (Expansion)  approach \cite{Cifuentes}, Moody's correlated
Binomial model \cite{Witt}, Long-range Ising model \cite{Molins}
are in this category.
The default correlation is defined by the 
simultaneous default probability. If we denote the probability
distribution of two asset $i,j$ as $P_{ij}
(x_{i},x_{j})$, the 
default correlation $\rho_{d}$ is defined by
\begin{equation}
\rho_{d}=\frac{P_{ij}(1,1)-P_{i}(1)\times P_{j}(1)}
{\sqrt{P_{i}(1)(1-P_{i}(1))}\sqrt{P_{j}(1)(1-P_{j}(1))}}. 
\label{d-a}
\end{equation}
Here, $P_{i}(x_{i})$ and $P_{j}(x_{j})$ are the probability
 distributions of $x_{i}$ and $x_{j}$ and they are calculated from
the joint probability distribution $P_{ij}(x_{i},x_{j})$.
  
In the second category, the models adopt a continuous variable for 
the earning rate of an asset and correlation between the earning rates
is introduced
\cite{Merton,Martin,Finger,Li}. 
On the assumption that the 
earing rates obey multivariate normal distribution with correlation 
$\rho_a$, the probability $P_{i,j}$ for the simultaneous default of 
the $i$-th and $j$-th assets is given by
\begin{equation}
P_{i,j}=
\frac{1}{2\pi \sqrt{1-\rho_{a}^{2}}} 
\int_{-\infty}^{\Phi^{-1}(P_{i})}
\int_{-\infty}^{\Phi^{-1}(P_{j})}\exp(-\frac{u^{2}-2\rho_{a}uv+v^{2}}
{2(1-\rho_{a}^{2})})dudv  .
\end{equation}
Here $P_{i}$ and $P_{j}$ are the default probabilities of the
$i$-th and $j$-th assets and $\Phi^{-1}(x)$ is the inverse function of the
normal distribution function. The variables $u,v$ mean the earing rates
of the two assets. If the random variable $u$ (or $v$) becomes lower than 
$\Phi^{-1}(P_{i})$ (resp. $\Phi^{-1}(P_{j})$), 
the $i$-th ($j$-th) asset  is judged to be defaulted.
The correlation parameter $\rho_{a}$ is named as ``asset correlation'' 
and $\rho_{a}$ and $\rho_{d}$ are related via the equation (\ref{d-a}).
The conditionally independent model \cite{Martin}, 
such as the Merton based model, the credit metrix model \cite{Finger}
and the copula model \cite{Li} are in 
the second category.

The reason why default or asset correlations are  widely discussed
recently is that the pricing of  Asset backed Security (ABS), 
like CDO, needs detailed information about 
the probabilities $P(N_{d})$ for $N_{d}$ defaults.
Here CDO is  an abbreviation for Collateralised debt obligation, 
which is a financial innovation to securitise portfolios of
defaultable assets.  The portfolio of the underlying debts (assets)
 collateralizes the securitites (obligations), CDO is  a kind of ABS.
Securitization by CDO, we mean to synthesize securities
 based on a pool of many assets, like loans (CLO), commercial bonds
 (CBO) etc. 
In the process, layered structure is introduced and  securities with 
high priority (reliability), which is called senior tranche, and 
those with low priority (called subordinated tranche or equity) are
synthesized.
Between the senior tranche and the equity, the mezzanine tranche with middle 
priority is also synthesized.
The difference between them is that if some of the assets 
in the pool are defaulted, the security with lower priority 
loses its value at first. If the rate of defaulted assets $\frac{N_{d}}{N}$ 
exceeds some threshold value $P_{c}$, e.g. $P_{c}=7\%$ for the mezzanie
tranche and $P_{c}=10\%$ for the senior tranche,  
those with higher priority begin to lose their values.
The equity play the role of ``shock absorber''.
By the ``tranche'' structure, the risk of the senior tranche is reduced
 and investors feel safe about the investment. 
On the other hand, the interest rates of the securities of the equity and
the mezzanine tranche 
are set to be higher than those of the senior tranche and  
the subordinated tranches are  high-risk-high-return products.

The default correlation becomes important when one try to estimate
the expected loss in each tranche, which is essential in the evaluation
of its price (premium). 
 For example, we assume a pool of $N$ homogeneous assets 
with  default probability
$P_{d}$. If there is no
correlation between the defaults, $P(N_{d})$ is the binomial
distribution $B(N,P_{d})$ and has a peak at $N_{d}=P_{d} N$.
The standard deviation is $\sqrt{N P_{d}(1-P_{d})}\simeq \sqrt{N P_{d}}$
for small $P_{d}$. If the threshold value $P_{c}$
 is large enough, the upper tranche does not suffer from the defaults
in the pool. On the other hand, in the extreme case where 
the default correlation is $\rho_{d}=1$, all assets behave 
in the same way and there are only two cases. 
One case is that all assets are not defaulted and the probability
for the case is $1-P_{d}$. The other case is that all assets are
defaulted simultaneously and the probability is $P_{d}$.
In the strong correlation limit ($\rho_{d}=1$), 
when there occur defaults, all assets become defaulted simultaneously.
Both senior and subordinated tranches lose their values completely.
If there occurs no default, both tranche does not suffer from any
damage.
The essential problem is to know the $\rho_{d}$ dependence of 
the probabilities $P(N_{d})$. It is important to estimate  
the expected loss rate of  each tranche based on $P(N_{d})$.
In addition, we should  also study   
which probabilistic model is good or useful in order to describe
 the behaviors of the assets.
 
This paper deals with these problems. 
The organization of the paper is as 
follows. In \sref{model}, we study the $(J,H)$ phase diagram of
finite size long-range Ising model and
show
that the 
assets begin to be  correlated in the ``Two Peak'' Phase 
in the $(J,H)$ plane.
 The realistic magnitude of the default correlation ranges from
1 \%  to  several \% \cite{Schonbucher}, 
only the Two Peak Phase
is interesting from the financial engineering viewpoint.
\Sref{theory} is devoted to the calculation of the
important parameter $P_{d}$ and $\rho_{d}$ in terms of $J,H$ and $N$.
Here, we develop a perturbation method which is based on the discussions
in \sref{model}. Up to zero-th order in the perturbation theory, 
$P(N_{d})$ is expressed as the superposition
of two binomial distributions, corresponding to the two peaks of $P(N_{d})$.
The developed  method and obtained relations 
are useful when one apply the
long-range Ising model to the evaluation and hedging of the securities
with tranches. 
In \sref{evaluation}, we study the $\rho_{d}$
dependence
of $P(N_{d},\rho_{d})$ 
and of the expected loss rates of the tranches.
For the latter purpose, 
we introduce the cumulative distribution $D(i,\rho_{d})$ 
and discuss that they are directly related  
with the average loss rates  of tranches.
As the correlation becomes strong (with fixed default probability
$P_{d}$), the left peak becomes taller and moves towards to the 
origin ($N_{d}=0$).
The right peak also becomes taller and shifts to $N_{d}=N$.
 Its area approaches to $P_{d}$ as $\rho_{d}$ comes close to 1. 
These behaviors are different from those of the binomial expansion
approach, where $P(N_{d})$ has only one peak and its shape
becomes broader as $\rho_{d}$ increases. We then discuss 
the $\rho_{d}$ dependence of $D(i,\rho_{d})$. 
 $D(i)$ for large $i$ 
increases linearly with $\rho_{d}$ and the senior tranche
cannot avoid  the  default damage of the assets pool, even when we set
$P_{c}$ to be large. This crucial behavior of the long-range Ising model
 has been pointed out previously \cite{Molins}, we have clarified 
the importance in the evaluation of the tranches. 
\Sref{conclusion} is dedicated to concluding remarks and 
future problems. We discuss the usefulness of the long-range Ising model
 from the viewpoint of financial engineering.

\section{Model and Phase Diagram in $(J,H)$ plane}
\label{model}

We use Ising Spin variables $S_{1},S_{2},\cdots,S_{N}=\pm 1$ which  
represent states of assets in the reference pool.
Here $S_{i}=-1$ indicates 
default of $i$-th asset and $S_{i}=1$ means that the $i$-th 
asset is not defaulted.
We denote the number of $S=\pm 1$ spins by $N_{\pm}$, so the number of 
defaulted assets $N_{d}$ is $N_{-}$. 
The probability distribution for the states of the assets 
is assumed to be described by the following canonical distribution with the long-range
Ising model of a finite system with $N$ spins and 
the exchange interaction $\frac{J}{N}$ and 
the external filed $H$, which are measured in units of Boltzmann
constant times temperature.
\begin{equation}
P(S_{1},S_{2},\cdots,S_{N})=
\frac{1}{Z_{N}(J,H)}
\exp \left( \frac{J}{2N} 
 \sum_{1\le i,j \le N}S_{i}S_{j}+H\sum_{i=1}^{N}S_{i}
\right).
\label{Ising}
\end{equation}
We do not omit  the $i=j$ terms in the Hamiltonian for later convenience.
As is well-known, the exchange interaction 
$-\frac{J}{N} S_{i}S_{j}$ controls the strength of the
correlation between $S_{i}$ and $S_{j}$ and  the external field $H$ favors
one of the two spin states. In the actual case where 
the spin variable represents
the states of the assets, the 
default probability $P_{d}$ is at most a few percent and 
almost all assets are not defaulted ($S=1$). The sign of the external
field $H$ is set to be $H>0$. 

The reason to choose the long-range Ising
model is that it gives the default distribution $P(N_{d})$ directly.
In \cite{Molins}, another motivation for the long-range Ising model has
 been discussed and their conclusion is that 
the model is the most natural choice from the viewpoint of 
the Maximum 
Statistical Entropy principle.
The two parameters $J$ and $H$ are introduced as Lagrange multipliers
which  ensure that the default probability and the default correlation
 of the model are $p_{d}$ and $\rho_{d}$.
From the economical viewpoint, we can interpret the model as a kind
of factor  model. Here, the term 'factor' means the systematic risk
factor 
or the state of the business cycle \cite{Schonbucher}.
In a boom, we have fewer defaults than in a recession.
We denote the state of the business cycle as $H'$  
and assume that
the defaults of the assets are  independent from each other, 
conditional on the realization of the systematic factor $H'$.
The joint probabilities for the assets $S_{1},S_{2},\cdots,S_{N}$
and the business cycle variable $H'$ is assumed to be written as
\begin{equation}
P_{factor}(S_{1},S_{2},\cdots,S_{N},H')=\frac{1}{Z_{N}(J)}
\exp \left( H' \sum_{i=1}^{N}S_{i}
\right)\times P(H').
\end{equation}
Here, the random variable $H'$ obeys the probability density function $P(H')$
 and the denominator $Z_{N}(J)$ is the normalization term.
Condition on the realization $H'=H$, the each asset state becomes independent
from each other and the default probability $P_{d}$ is given as
\begin{equation}
P_{d}=\mbox{Prob}(S_{i}=-1)=\frac{e^{-2H}}{1+e^{-2H}}.
\end{equation}
The default probability $P_{d}$ is a decreasing function of $H$
 and $H$ for a boom (recession) is large (small).
In order to derive the long-range Ising model starting from the
above factor mode, we assume that $H'$ obeys the standard normal
distribution with mean $H$ and variance $J/N$.
\begin{equation}
P(H')=\frac{1}{\sqrt{2\pi J}}\exp(-\frac{(H'-H)^{2}}{2J/N})
\end{equation}
By averaging  over the possible realization of $H'$ weighted with 
the above $P(H')$, we obtain the expression for the long-range Ising
model. 
\begin{eqnarray}
&&\int_{-\infty}^{\infty}P_{factor}(S_{1},S_{2},\cdots,S_{N},H')dH'
=\int_{-\infty}^{\infty} \frac{1}{Z_{N}(J)}
\exp \left( H' \sum_{i=1}^{N}S_{i}
\right)\times P(H')dH' \nonumber \\
&&=\frac{1}{Z_{N}(J,H)}
\exp \left( \frac{J}{2N} 
 \sum_{1\le i,j \le N}S_{i}S_{j}+H\sum_{i=1}^{N}S_{i}
\right).
\end{eqnarray}
The validity of the Maximum 
Statistical Entropy principle or the factor model with the normally
distributed business factor $H'$ should be checked by the 
comparison with other  more reliable models.

\begin{figure}[htbp]
\begin{center}
\includegraphics[width=0.7\linewidth]{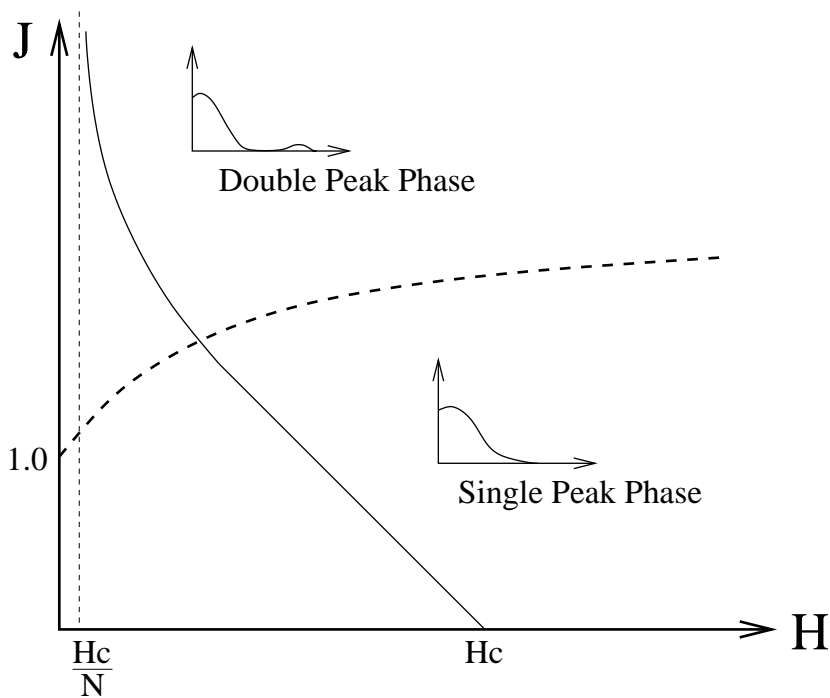}
\end{center}
\caption{Phase diagram in $(J,H)$ plane. For large $H$ and small $J$, 
$P(N_{d})$ has a single peak at $N_{d} \simeq P_{d} N$. 
We call the
 region as ``Single Peak'' Phase. 
For small $H$ and large $J$, there are
 two
peaks in $P(N_{d})$ and we call the region ``Two Peak'' Phase.  
The phase boundary is depicted with the broken line (\broken).
The solid
 line (\full) corresponds to a constant $P_{d}$ line. 
It starts at $(J,H)=(0,H_{C})
$, where $\rho_{d}=0$. In the $J \to \infty$ limit, the line
approaches $H=\frac{H_{c}}{N}$ asymptotically and 
$\rho_{d} \to 1$.}
\label{fig:phase}
\end{figure}

The Hamiltonian of the long-range Ising model depends on the spin 
variables only through the combination of the magnetization 
$M=\sum_{i=1}^{N}S_{i}$. There is a simple relation between
$N_{-}=N_{d}$ and $M$ as $M=N_{+}-N_{-}=N-2 N_{d}$,
the default number distribution function $P(N_{d})$ is 
\begin{equation}
P(N_{d})=\frac{\exp\left(\frac{J}{2N}N^{2}+HN \right)}{Z_{N}(J,H)}
{}_{N}C_{N_{d}}
\exp\left(\frac{2J}{N}N_{d}^{2}-(2J+2H)N_{d}   \right).
\label{P1}
\end{equation}
The default probability $P_{d}$ is defined by the expectation value of
$N_{d}$ as
\begin{equation}
P_{d} \equiv \frac{<N_{d}>}{N}.
\end{equation}
Here $<\hspace*{0.4cm} >$ is the expectation value with the probability
distribution (\ref{P1}). 
For $J=0$, the probability distribution (\ref{P1}) becomes
 that of the binomial distribution $B(N,P_{d})$ and there is a 
relation between $H$ and $P_{d}$ as
\begin{equation}
H=-\frac{1}{2}\log \left( \frac{P_{d}}{1-P_{d}}\right)  \hspace*{0.3cm
}\mbox{for}\hspace*{0.3cm} J=0.
\end{equation} 
We denote this value of $H$ as $H_{C}$.
On the other hand, for $J \to +\infty$ limit, there are only 2
configurations with all spins up or all spins down that have nonzero
 probabilities. The probabilities are
\begin{equation}
P(1,1,\cdots,1)=\frac{1}{1+e^{-2HN}}
\hspace*{0.5cm} P(-1,-1,\cdots,-1)=\frac{e^{-2HN}}{1+e^{-2HN}}. \label{P0}
\end{equation}
From the relation $P(-1,-1,\cdots,-1)=P_{d}$, one obtains 
the following relation between $H$ and $P_{d}$ for $J=\infty$ as 
\[
H_{J=\infty}=\frac{H_{C}}{N}. 
\]
For general $J,H$ and $N$, it is difficult to obtain $P_{d}$. 
However, for large enough $N$, by changing variable from $N_{-}$
to $n_{-}=\frac{N_{-}}{N}$ in eq.(\ref{P1}), we can estimate
$P_{d}=<n_{-}>$ by the saddle point approximation.
The saddle point equation is  
\begin{equation}
n_{-}=\frac{e^{4J n_{-}-(2J+2H)}}{e^{4J n_{-}-(2J+2H)}+1}. \label{P2}
\end{equation}
Of course, by changing variable from $n_{-}$ to 
the magnetization per spin $m=1-2n_{-}$, 
the saddle point equation is transformed into the 
famous self-consistent equation of the magnetization
$m=\tanh(J m+H)$ \cite{Stanley}. 
Depending on the values of the parameters $J,H(>0)$,
 there are two cases. For large $H$ and small $J$, the equation
 (\ref{P2})  has only one solution $n_{-*}$. 
We call this region in 
the $(J,H)$ plain as ``One Peak'' Phase, because the probability 
distribution $P(N_{d})$ has a single peak at $N  n_{-*}$.
$P_{d}$ is almost the same with $n_{-*}$ in the One Peak Phase.   
For small $H$ and large $J$, the equation (\ref{P2}) has 
three solutions, two are 
at maxima $n_{-1*}< n_{-2*}$ and one is at minimum. 
We call the   
region in the $(J,H)$ plane as ``Two Peak'' Phase, 
as the reader may easily anticipate the reason.  
 In the case, there is no simple relation between $P_{d}$ and the
 solutions $n_{-1*},n_{-*2}$. If $H>0$ is large, the solution $n_{-1*}$
is almost the same with $P_{d}$. However, when the
correlation $\rho_{d}$ is large, the strength of $H$ is of the order of 
$\Or(\frac{1}{N})$ and we cannot neglect the second peak $n_{-*2}$.
In the case, $n_{-1*}<P_{d}$ and the average value of $n_{-1*}$ 
and $n_{-2*}$ with $P(n_{-*1})$ and $P(n_{-*2})$
corresponds to the value of $P_{d}$. 
For example, when $\rho_{d}=1$ and $J=\infty$, the 
average value of $n_{-1*}=0$ and $n_{-2*}=1$ 
with probabilities eq.(\ref{P0}) is equal  to $P_{d}$.
In \fref{fig:phase}, we summarize the situation. 
The solid curve (\full) in the $(J,H)$
plane corresponds to the constant $P_{d}$ line. The dotted line (\broken)
is the ``phase transition'' line between the One-Peak Phase and the Two
Peak Phase. In the remainder of the
section, we study the correlation $\rho_{d}$ in the $(J,H)$ phase
diagram. We will see that $\rho_{d}$  is almost zero in the One Peak Phase.
Only in the Two Peak Phase $\rho_{d}$ can take nonzero value.

\begin{figure}[htbp]
\begin{center}
\includegraphics[width=0.7\linewidth]{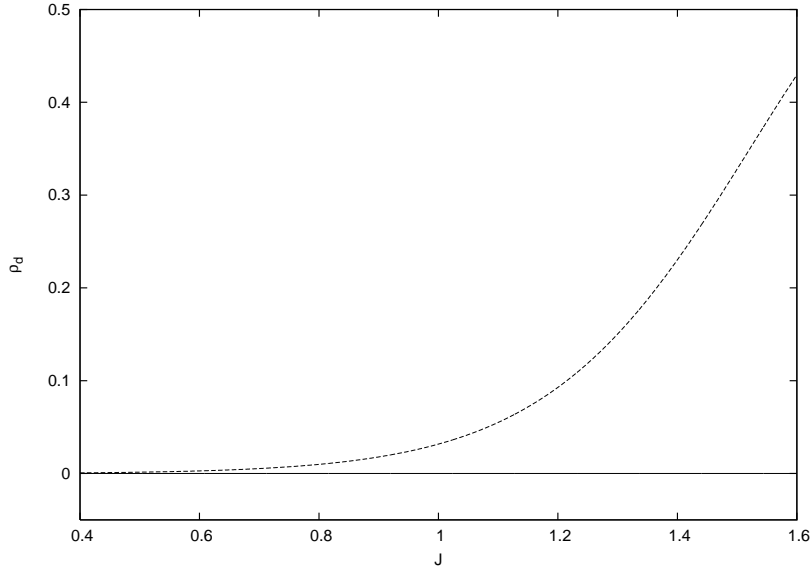}
\end{center}
\caption{Second peak contribution in $\rho_{d}$.
Along the line with eq.(\ref{mean_a}), we plot the approximated
estimation for $\rho_{d}$. The lower solid line (\full) shows the data 
from eq.(\ref{ap1}) and the data from eq.(\ref{ap2}) are depicted 
with upper dotted curve (\dotted).  
}
\label{fig:saddle12}
\end{figure}

We discuss the default correlation $\rho_{d}$ and 
recall the definition $(\ref{d-a})$. In order to obtain
$P(S_{1},S_{2})$, in equation (\ref{Ising}), we
take the trace over $S_{3},S_{4},\cdots,S_{N}$.
\begin{equation}
P(S_{1},S_{2})=\prod_{j=3}^{N}\sum_{S_{j}=\pm 1} 
P(S_{1},S_{2},\cdots,S_{N}).
\end{equation}
The trace over $S_{3},\cdots,S_{N}$ is replaced by the summation
over $N_{-}$ of $N''=N-2$ Spins. We obtain
\begin{eqnarray}
&&P(S_{1},S_{2})=\frac{1}{Z_{N}(J,H)}
\exp\left(\frac{J}{N}S_{1}S_{2}+H(S_{1}+S_{2})  \right) \times
\nonumber 
\\
&& \sum_{N_{-}=0}^{N''}{}_{N''}C_{N_{-}}
e^{(\frac{J}{N}(S_{1}+S_{2})(N''-2N_{-})+
\frac{J}{N}(N''-2N_{-})^{2}+H(N''-2N_{-}) 
)}.
\end{eqnarray}
If the system size $N$ is large, 
the summation over $N_{-}$ is replaced by the estimation at the 
saddle points. 
In the One-Peak region, the saddle point 
is at $n_{-*}=P_{d}$ and $P(S_{1},S_{2})$ is given by
\begin{equation}
P(S_{1},S_{2})\propto 
\exp(\frac{J}{N}S_{1}S_{2}+H(S_{1}+S_{2}))\exp(-J(S_{1}+S_{2})
\times 2 n_{-*}). \label{ap1}
\end{equation}
In the Two Peak region, two saddle points contribute to the
summation and $P(S_{1},S_{2})$ is estimated as
\begin{eqnarray}
&&P(S_{1},S_{2})\propto 
\exp(\frac{J}{N}S_{1}S_{2}+H(S_{1}+S_{2}))
\times  \nonumber \\
&& [ e^{-J(S_{1}+S_{2})
\times 2 n_{-1*}}\times  P(n_{-1*})+
     e^{-J(S_{1}+S_{2})
\times 2 n_{-2*}} \times P(n_{-2*}).
] \label{ap2}
\end{eqnarray}
Here $P(n_{-1*})$ and $P(n_{-2*})$ are the probabilities
for the two peaks $n_{-1*}$ and $n_{-2*}$.
In the One Peak phase, the constant $P_{d}$ line in the
$(J,H)$ plane
is almost given by the following relation between $J$ and $H$ 
\begin{equation}
H=\tanh^{-1}(1-2 P_{d})-J(1-2P_{d}). \label{mean_a}
\end{equation}
We calculate the default correlation $\rho_{d}$ with equations
(\ref{ap1}) and (\ref{ap2})
on the above approximate constant $P_{d}$ line.
About the two saddle points $n_{-1*},n_{-2*}$ and their probabilities
$P(n_{-1*}),P(n{-2*})$, we take them the values 
at $J=\infty$ and $\rho_{d}=1$. We set $n_{-1*}=0,n_{-2*}=1$ and 
$P(n_{-1*})=1-P_{d},P(n_{-2*})=P_{d}$.
We set $P_{d}=0.01$ and we plot $\rho_{d}$ vs $J$ in \fref{fig:saddle12}.
We see that the correlation with equation (\ref{ap1}), 
which is plotted with solid line (\full),
does not become large 
even in the Two Peak Region. On the other hand, 
$\rho_{d}$ with equation
(\ref{ap2}), which is depicted with dotted line (\dotted) 
becomes large in the Two Peak Region.
We see that the existence of the second peak in  $P(N_{d})$ 
plays a crucial role in the emergence of correlation 
in the long-range Ising model.

\section{Perturbative Calculation and Second Peak Contribution}
\label{theory}

In this section, we try to calculate several quantities
of interest of the long-range Ising model. In particular,
we obtain the expressions for $P_{d}$ and $\rho_{d}$ in terms of the 
model parameter $J,H$ and $N$. 
In addition, we also obtain the
expression for the 
probability (or weight) of the second peak $P_{all}$, 
which means that almost all assets are defaulted \cite{Molins}.
The probability $P_{all}$ plays a crucial
role when one discuss the evaluation 
of the tranche.

When one calculate $P_{d},\rho_{d}$, one way is to calculate
$<S_{i}>$ and $<S_{i}S_{j}>$. Here, we calculate the moment of $N_{d}$
 with the probability distribution eq.(\ref{P1}). The default probability
is then given by $P_{d}=\frac{<N_{d}>}{N}$. About the default
correlation $\rho_{d}$, we start from the following relation.
\begin{equation}
\rho_{d}=\frac{P(-1,-1)-P(-1)^{2}}{P(-1)(1-P(-1))}=
\frac{<S_{1}S_{2}>-<S_{1}><S_{2}>}{(1-<S_{1}>)(1+<S_{1}>)}
\end{equation}
The magnetization $M=N<S_{1}>$ and $N_{d}$ is related as $M=N-2N_{d}$ and
$N(N-1)<S_{1}S_{2}>=<M^{2}>-N$, we obtain the following expression
\begin{equation}
\rho_{d}=\frac{\sigma_{N_{d}}^{2}+\frac{1}{N-1}(<N_{d}^{2}>-N<N_{d}>)}
{<N_{d}>(N-<N_{d}>)}.   \label{rho_d}
\end{equation}
In order to calculate the moment $<N_{d}^{l}>$ with eq.\eref{P1}, the
quadratic term $\exp(\frac{2J}{N}N_{d}^{2})$
prevents us from  
taking  summation over $N_{d}$. As we have noted previously, the
distribution with $J=0$ is binomial distribution and taking summation
over $N_{d}$ is easy.  
In addition, the $P_{d}$ is at most a few percent
and the distribution $P(N_{d})$ have a peak very close to $N_{d}=0$ 
(and the second peak at $N_{d} \simeq N$ in the Two Peak Phase). 
We expand the quadratic term as
\begin{equation}
\exp(\frac{2J}{N}N_{d}^{2})=\sum_{k=0}^{\infty}\frac{1}{k!}
(\frac{2J}{N}N_{d}^{2})^{k}.   \label{exp}
\end{equation}
and perform the calculation of the moment $<N_{d}^{l}>$ 
perturbatively.  The expansion is about $\frac{2J}{N}N_{d}^{2}$, which
is evaluated as 
$\frac{2J}{N}N_{d}^{2}\simeq 2J N P_{d}^{2}$. In the actual risk
portfolio problem, $P_{d}$ is at most $2 \sim 3 \%$ and the system size
$N$ is several hundred, the
perturbative approximation is considered to be applicable. 
We also note that, in the Two Peak Phase, the above expansion 
should be carried out also at $N_{d}=N$. 

In order to perform the calculation in more concrete manner,  
 we use variables $N_{\pm}$ and start from the
following expression for the Hamiltonian.
\begin{eqnarray}
-{\cal H}&=&\frac{J}{2N}M^{2}+HM \nonumber \\
&=&\frac{J}{2N}N^{2}+HN-(2H+2J)N_{-}+\frac{2J}{N}N_{-}^{2}  \label{e1}\\
&=&\frac{J}{2N}N^{2}-HN-(2J-2H)N_{+}+\frac{2J}{N}N_{+}^{2}.   \label{e2}
\end{eqnarray}
In the vicinity of $N_{-}=0$, we denote $P(N_{-})$ as $P_{-}(N_{-})$
and we can expand the quadratic term in
eq.(\ref{e1}). Likewise, in the vicinity of $N_{-}=N (N_{+}=0)$, 
we call $P(N_{-})$ as $P_{+}(N_{-})$ and it can also be expanded in $N_{+}$. 
\begin{eqnarray}
P_{-}(N_{-}) &=
\frac{1}{Z}
 {}_{N}C_{N_{-}}
e^{HN}e^{
-(2H+2J)N_{-}+\frac{2J}{N}N_{-}^{2}} \nonumber 
\\
&=\frac{1}{Z}
 {}_{N}C_{N_{-}}
e^{HN}e^{
-(2H+2J)N_{-}}
\times \sum_{k=0}^{\infty}\frac{1}{k!}
(\frac{2J}{N}N_{-}^{2})^{k}
 \\
P_{+}(N_{-})&= 
\frac{1}{Z}
 {}_{N}C_{N-N_{-}}
e^{-HN}
e^{-(2J-2H)(N-N_{-})
+\frac{2J}{N}(N-N_{-})^{2}}
  \nonumber \\
&=
\frac{1}{Z}
 {}_{N}C_{N-N_{-}}
e^{-HN}
e^{-(2J-2H)(N-N_{-})} \times
\sum_{k=0}^{\infty}\frac{1}{k!}
(\frac{2J}{N}(N-N_{-})^{2})^{k}.
\end{eqnarray}
$Z$ is the normalization constant to ensure that
$\sum_{N_{-}=0}^{N}P(N_{-})=1$. To the zero-th order perturbation 
approximation $P_{-}(N_{-})$ and $P_{+}(N_{-})$ are binomial 
distributions and
$P(N_{-})$ is given by the superposition of these distributions.
We summarize the 
situation as
\begin{equation}
P(N_{-})=
\cases{
P_{-}(N_{-}) &   $(0 \le N_{-} \le L)$  \\
P_{+}(N_{-}) &   $(0 \le N-N_{-} < N-L)$.  \\ 
}
\end{equation}
Here $L$ is set to be at the middle of the interval $[0,N]$.

The moment $<N_{-}^{l}>$ is calculated with the following equation.
\begin{equation}
<N_{-}^{l}>=\sum_{N_{-}=0}^{N}P(N_{-})N_{-}^{l}
=\sum_{N_{-}=0}^{L}P_{-}(N_{-})N_{-}^{l}
+\sum_{N_{-}=L}^{N}P_{+}(N_{-})N_{-}^{l}. \label{moment}
\end{equation}
The summation over $N_{-}$ is from $0$ to $L$, however $P_{-}(N_{-})$
 damps rapidly in $N_{-}$, it is not so bad to change the
range from $[0,L]$ to $[0,N]$. About $P_{+}(N_{-})$ the range of $N_{-}$
is $[L,N]$. We change variable from $N_{-}$ to $N_{+}=N-N_{-}$ and 
denote the probability distribution 
$P_{+}(N_{-}=N-N_{+})$ also as $P_{+}(N_{+})$. 
\begin{equation}
P_{+}(N_{+})=
\frac{1}{Z}
 {}_{N}C_{N_{+}}
e^{-HN}
e^{-(2J-2H)(N_{+})+\frac{2J}{N}(N_{+})^{2}}.  
\end{equation}
$P_{+}(N_{+})$ also damps
 rapidly in $N_{+}$, we will change the summation 
range from $[0,N-L)$ to $[0,N]$.
$<N_{-}^{l}>$ is then calculated perturbatively as
\begin{eqnarray}
&<N_{-}^{l}>=\sum_{N_{-}=0}^{N}P_{-}(N_{-})N_{-}^{l}
+\sum_{N_{+}=0}^{N}P_{+}(N_{+})(N-N_{+})^{l} \nonumber \\
&=
\frac{1}{Z}e^{HN} \sum_{N_{-}=0}^{N}
 {}_{N}C_{N_{-}}e^{-(2H+2J)N_{-}}
\sum_{k=0}^{\infty}
\frac{1}{k!}(\frac{2J}{N}N_{-}^{2})^{k} N_{-}^{l} \nonumber \\
&+\frac{1}{Z}e^{-HN} \sum_{N_{+}=0}^{N}
 {}_{N}C_{N_{+}}e^{-(2J-2H)N_{+}}
\sum_{k=0}^{\infty}
\frac{1}{k!}(\frac{2J}{N}N_{+}^{2})^{k} (N-N_{+})^{l} .
\end{eqnarray}
The normalization constant $Z$ is calculated as
\begin{eqnarray}
Z&=&e^{HN} \sum_{N_{-}=0}^{N}
 {}_{N}C_{N_{-}}e^{-(2H+2J)N_{-}}
\sum_{k=0}^{\infty}
\frac{1}{k!}(\frac{2J}{N}N_{-}^{2})^{k}  \nonumber \\
&+& e^{-HN} \sum_{N_{+}=0}^{N}
 {}_{N}C_{N_{+}}e^{-(2J-2H)N_{+}}
\sum_{k=0}^{\infty}
\frac{1}{k!}(\frac{2J}{N}N_{+}^{2})^{k} \nonumber \\
&=&Z_{-}+Z_{+}.
\label{Z}
\end{eqnarray}
In equation (\ref{Z}), we denote the two terms as $Z_{\pm}$, which come
from the summation over $N_{-}$ and $N_{+}$.

\begin{figure}[htbp]
\begin{center}
\includegraphics[width=0.7\linewidth]{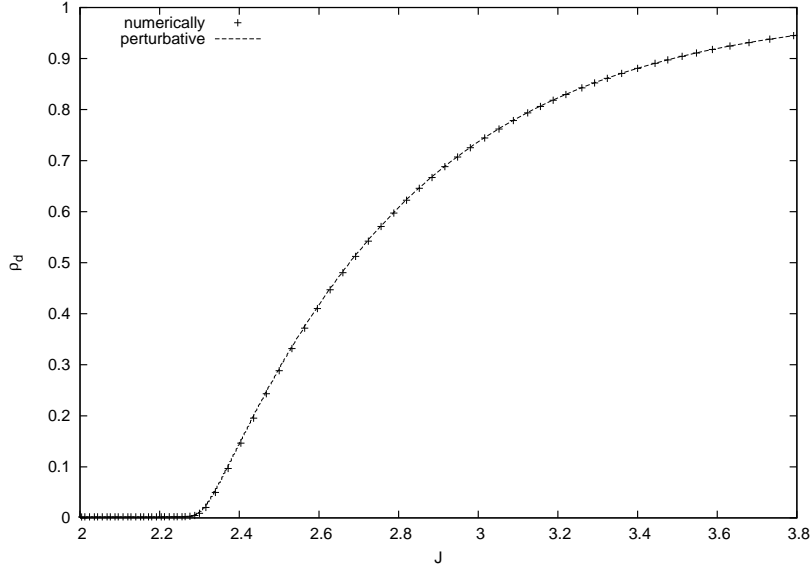}
\end{center}
\caption{The default correlation $\rho_{d}$ on a constant $P_{d}$ line.
$P_{d}=0.01$ and we plot $\rho_{d}$ versus $J$. $H$ is set to be the
 value which realize $P_{d}=0.01$ for $N=50$. 
The solid line (\full) shows
the result from the perturbative calculation up to second order in 
$\frac{2J}{N}$ and the line with $+$ symbols 
shows numerically calculated exact data.}
\label{fig:hikaku1}
\end{figure}

In the above calculation, moments of the binomial distribution
appears frequently. We introduce the following unnormalized 
binomial moments $\left[ N_{\pm} ^l \right]$.
\begin{equation}
\left[ N_{\pm} ^l \right] 
=\sum_{N_{\pm} = 0}^{N} {}_{N}C_{N_{\pm}} e^{\beta_{\pm}
N_{\pm}} N_{\pm}^{l}.
\end{equation}
The parameters $\beta_{\pm}$ are defined as $\beta_{\pm}=-2(J \mp H)$.
Calculations of $\left[ N_{\pm} ^l \right]$ is straightforward. 
The zero-th moment $[1_{\pm}]$ is given by   
\begin{equation}
 \left[ 1_{\pm} \right]=\left( 1 + e^{\beta_{\pm}}\right) ^{N}.
\label{eq:0th moment}
\end{equation}
The  $l$-th moment $\left[ N_{\pm} ^l \right]$ is then obtained by
differentiating $[1_{\pm}]$ with respect to $\beta_{\pm}$ repeatedly.
\begin{equation}
\left[ N_{\pm}^{l}\right] = \frac{\partial^{l}}{\partial \beta_{\pm} ^{l}}
\left( 1 + e^{\beta_{\pm}}\right)^{N}.
\end{equation}
We show the results for the first 6 moments, which are necessary for 
the second order perturbative calculation. 
\begin{eqnarray}
 \left[ N_{\pm}^{1}\right] = & \left( 1 + e^{\beta_{\pm}} \right)^{N}
  (N_{1}x_{\pm})    \\
 \left[ N_{\pm}^{2}\right] = & \left( 1 + e^{\beta_{\pm}} \right)^{N}
  (N_{1}x_{\pm} + N_{2}x_{\pm}^{2})   \\
 \left[ N_{\pm}^{3}\right] = & \left( 1 + e^{\beta_{\pm}} \right)^{N}
  (N_{1}x_{\pm} + 3N_{2}x_{\pm}^{2} + N_{3}x_{\pm}^{3}) \\
 \left[ N_{\pm}^{4}\right] = & \left( 1 + e^{\beta_{\pm}} \right)^{N}
  (N_{1}x_{\pm} + 7N_{2}x_{\pm}^{2} + 6N_{3} x_{\pm}^{3}+
  N_{4}x_{\pm}^{4}) \\
 \left[ N_{\pm}^{5}\right] = & \left( 1 + e^{\beta_{\pm}} \right)^{N}
  (N_{1}x_{\pm} + 15 N_{2}x_{\pm}^{2} + 25 N_{3} x_{\pm}^{3}+
  10 N_{4}x_{\pm}^{4} +N_{5}x_{\pm}^{5}) \\
 \left[ N_{\pm}^{6}\right] = & \left( 1 + e^{\beta_{\pm}} \right)^{N}
(N_{1}x_{\pm} + 31 N_{2}x_{\pm}^{2} 
\nonumber \\
 & \hspace*{2.5cm} + 90 N_{3} x_{\pm}^{3}+
  65 N_{4}x_{\pm}^{4} +15 N_{5}x_{\pm}^{5}+N_{6}x_{\pm}^{6}), 
\end{eqnarray}
where $N_{i} = \frac{N!}{\left(N -i \right) !}$ and $x_{\pm} =
\frac{e^{\beta_{\pm}}}{1+e^{\beta_{\pm}}}$.
In general, the $l$-th binomial moment $\left[ N_{\pm} ^l \right]$ is 
calculated as
\begin{equation}
\left[ N_{\pm}^{l}\right] = 
\left( 1 + e^{\beta_{\pm}} \right)^{N}
\sum_{k=1}^{l} y^{l}_{\pm k}N_{k}x_{\pm}^{k}, 
\end{equation}
where the coefficients $y^{l}_{\pm k}$ for $N_{k}x_{\pm}^{k}$ 
is calculated with the following recursive relations.
\begin{equation}
y^{l}_{\pm k} = y^{l-1}_{\pm k-1} + k y^{l-1}_{\pm k}
\end{equation}
and with the conditions $y^{l}_{\pm k}=0$ for $k>l$ and $y^{1}_{1}=1$.

With these preparations, we are ready to write down the results.
The perturbative calculation of the normalization
 constant $Z$ is given as
\begin{eqnarray} 
Z&=Z_{-}+Z_{+} \nonumber \\
&=
e^{HN}\sum_{k=0}^{\infty}\frac{1}{k!}
\left(\frac{2J}{N}\right)^{k}[N_{-}^{2k}]
+
e^{-HN}\sum_{k=0}^{\infty}\frac{1}{k!}
\left(\frac{2J}{N}\right)^{k}[N_{+}^{2k}] .
\end{eqnarray}
 The moment $<N_{-}^{l}>$ is given by
\begin{eqnarray}
& <N_{-}^{l}>=
\frac{1}{Z}
e^{HN}\sum_{k=0}^{\infty}\frac{1}{k!}
\left(\frac{2J}{N}\right)^{k}[N_{-}^{2k+l}] \nonumber \\
&+ \frac{1}{Z}
e^{-HN}\sum_{k=0}^{\infty}\frac{1}{k!}
\left(\frac{2J}{N}\right)^{k}
\sum_{m=0}^{l}{}_{l}C_{m}
(-1)^{m}N^{l-m}[N_{+}^{2k+m}]  .
\end{eqnarray}

\begin{figure}[htbp]
\begin{center}
\includegraphics[width=0.9\linewidth]{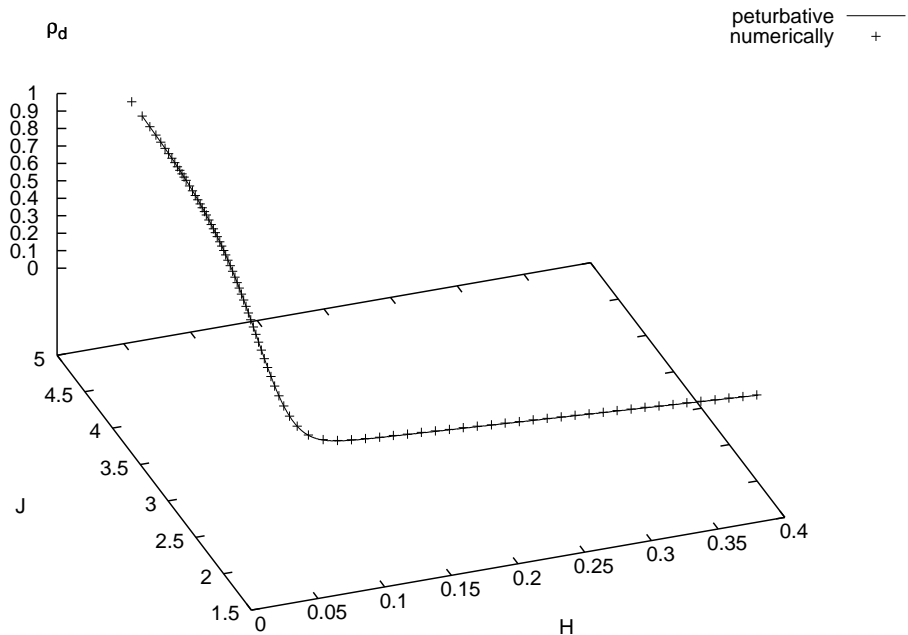}
\end{center}
\caption{3-dimensional plot of $\rho_{d}$ in $(J,H,\rho_{d})$ 
space. 
The solid line (\full) shows
 the result from the perturbative calculation up to second order in 
$\frac{2J}{N}$ and the line with $+$ symbols 
depicts numerically calculated exact data.
The conditions are the same with those 
in figure \ref{fig:hikaku1}.}
\label{fig:hikaku2}
\end{figure}

Putting these results into $P_{d}=\frac{<N_{d}>}{N}$ and
eq.(\ref{rho_d}), 
the expressions for $P_{d}$ and $\rho_{d}$ in
terms of model parameters $N,J,H$ are obtained. In addition, the weight
of the second peak $P_{all}$, that is the probability of almost 
all assets are defaulted, is estimated as 
\begin{equation}
P_{all}=\frac{Z_{+}}{Z}.  \label{pall}
\end{equation}
As we have noted previously, the zero-th order 
approximation means that we express $P(N_{d})$ as a 
superposition of two binomial distributions. In the case, the
 results for $Z$ and $P_{d},\rho_{d}$ can be written down
in the following simple expressions. 
\begin{eqnarray}
Z^{0}&=e^{HN}(1+e^{\beta_{-}})^{N}(1+\gamma^{N})
\\
P_{d}^{0}&=\frac{x_{-}+\gamma^{N}(1-x_{+})}{1+\gamma^{N}}
\\
\rho_{d}^{0}&=\frac{1}{P_{d}^{0}(1-P_{d}^{0})}
\frac{\gamma^{N}}{(1+\gamma^{N})^{2}}(1-(x_{-}+x_{+}))^{2} 
\\
\gamma&=e^{-2H}\left(\frac{1+e^{\beta_{+}}}{1+e^{\beta_{-}}}\right)
\end{eqnarray}
The subscript $(0)$ indicates the zero-th order 
perturbation results.
In \fref{fig:hikaku1}, we 
 shows the result for $\rho_{d}^{2}$ along the constant $P_{d}$ line.
$P_{d}$ is set to be $P_{d}=0.01$ and with solid line (\full) 
we show the data from the above
 perturbative calculation up to second order in $\frac{2J}{N}$.
The line with $+$ symbols 
depicts the numerical data.
 The two lines coincide well and the match is very good as long as $P_{d}$
is set to be small. \Fref{fig:hikaku2} is the 3-dimensional plot
of the data in $(J,H,\rho_{d})$ space. $\rho_{d}$ begins to be large
 in the Two Peak region and its rapid growth is well captured by the
above perturbative calculation.

\section{Effect of $\rho_{d}$ on $P(N_{d})$ and on average loss rates 
of tranches}
\label{evaluation}

We would like to discuss the effect of default correlation $\rho_{d}$
on the probabilities $P(N_{d})$ and 
on the tranche synthesized from the pool of the
homogeneous  assets. In order to discuss the latter case, 
we introduce the cumulative
distribution function $D(i,\rho_{d})$, which is directly related with
the average loss rate of the tranche.
The default rate $P_{d}$ and the system size $N$ is fixed.
When we show numerical data, we set $N=100$ and $P_{d}=0.05$.

\begin{figure}[htbp]
\begin{center}
\includegraphics[width=0.9\linewidth]{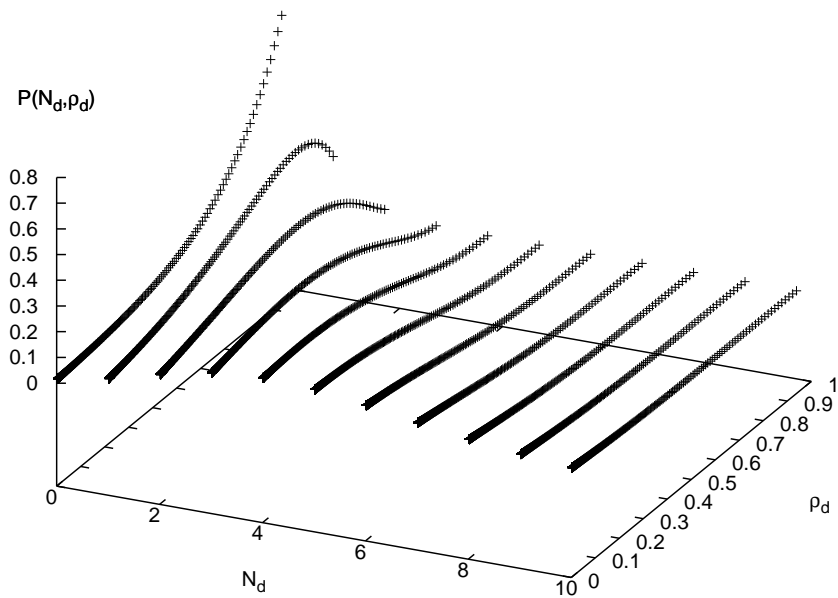}
\caption{
Plot of $P(N_{d},\rho_{d})$ vs $\rho_{d}$.
$0\le N_{d} \le  10$ and $N=100,P_{d}=0.05$.
}
\label{fig:0to10Pd}
\end{center}
\end{figure}

\begin{figure}[htbp]
\begin{center}
  \includegraphics[width=0.9\linewidth]{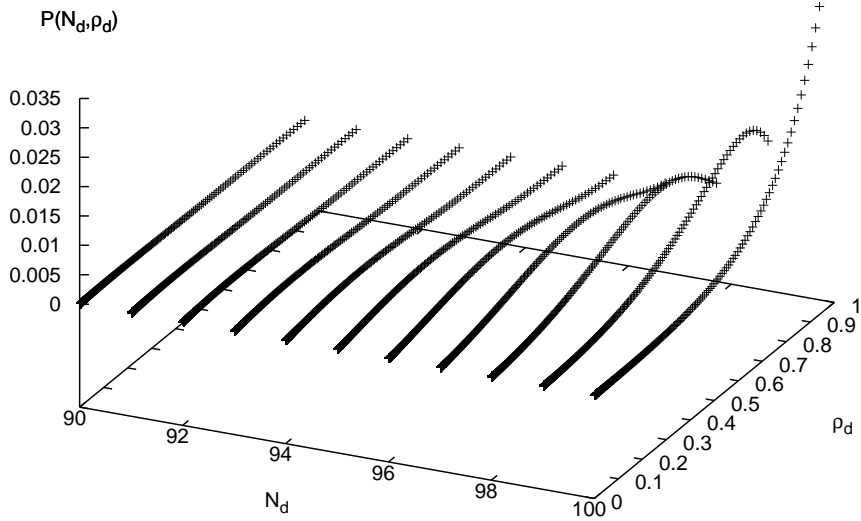}
  \caption{
Plot of $P(N_{d},\rho_{d})$ vs $\rho_{d}$.
$90 \le N_{d} \le  100$ and $N=100,P_{d}=0.05$.
}
\label{fig:90to100Pd}
\end{center}
\end{figure}

At first, we discuss the former case, the 
effect on the probabilities $P(N_{d},\rho_{d})$. Here we
write down their $\rho_{d}$ dependence explicitly.
The $\rho_{d}=0$ case is easy and $P(N_{d},0)$ is only the 
binomial distribution $B(N,P_{d})$. It has a single peak at
 $N P_{d}$ and the width is roughly $
2 \sqrt{N  P_{d}}$.
For $N_{d}=0$,
$P(0,0)=(1-P_{d})^{N}$ and as
$\rho_{d}$  becomes large, $P(0,\rho_{d})$ approaches $1-P_{d}$.
In \fref{fig:0to10Pd}, we plot $P(N_{d},\rho_{d})$ versus $\rho_{d}$ for
$N_{d}\le 10$. The system size $N=100$  and $P_{d}=0.05$.
$P(0,\rho_{d})$ grows monotonically as $\rho_d$ grows.
For $1 \le N_{d} \le 5=N P_{d}$, 
$P(N_{d},\rho_{d})$ at first increases and then decreases
as a function of $\rho_{d}$.
On the other hand, for $N_{d}\ge 5=N P_{d}$, 
$P(N_{d},\rho_{d})$ decreases 
with $\rho_{d}$.
$P_{d}$ is small and $P(N_{d},0)$ damps rapidly in $N_{d}$ 
for $N_{d} \ge 5$, 
$P(N_{d},\rho_{d})$ is almost zero 
for any $\rho_{d}$, which holds for $10 < N_{d}\le 90$. 
\Fref{fig:90to100Pd} depicts  the plots of 
$P(N_{d},\rho_{d})$ for $90 \le N_{d}\le 100$.    
$P(N,0)=P_{d}^{N}\simeq 0$ and $P(N,1)=P_{d}$, $P(100,\rho_{d})$
grows monotonically  from $0$ to $P_{d}=0.05$. For $90 \le N_{d} \le 99$,
 $P(N_{d},\rho_{d})$ is upward convex with respect to $\rho_{d}$.
The area of the second peak 
becomes greater with the increase of $\rho_{d}$ and $P(N_{d},\rho_{d})$ 
increases for $N_{d} \simeq N$. As $\rho_{d}$ becomes large, the width of the
second peak becomes narrow and $P(N_{d},\rho_{d})$ for $N_{d} \neq  N$
 decreases. On the other hand, $P(N,\rho_{d})$ increases
 monotonically to $P_{d}$.

To sum up, for $\rho_{d}=0$, $P(N_{d},0)$ is $B(N,P_{d})$ and 
it has a single peak. The width of the peak is order 
$2 \sqrt{N P_{d}}$  and it
is small for small $P_{d}$.  As $\rho_{d}$ grows, the system 
is in the Two Peak Phase.  
At the zero-th order perturbative approximation, $P(N_{-})
=P_{-}(N_{-})+P_{+}(N_{+})$
is a superposition of two binomial distributions.
$P_{-}(N_{d})$ has a peak at $N_{d} \le N  P_{d}$ and is
approximately obeys $B(N,x_{-})$. On the other hand, 
$P_{+}(N_{d})$ is $B(N,1-x_{+})$ and has a peak at $N_{d} \sim N$.  
The increase in $\rho_{d}$ accompanies the increase in  $J$, however  
the change of $H$  is not so large and it decreases slightly
(See \Fref{fig:phase}).
The first peak position of $P(N_{d})$, which is governed by 
$\beta_{-}=-2(J+H)$, moves towards to $N_{d}=0$ as $J$ increases.
The first peak 
becomes narrower and higher with the left slide and only $P(0,\rho_{d})$
grows monotonically. For $0 < N_{d} \le N  P_{d}$, 
$P(N_{d},\rho_{d})$ is upward convex with respect to $\rho_{d}$. 
$P(N_{d},\rho_{d})$ for $N_{d} > N  P_{d}$ damps with
$\rho_{d}$ monotonically. 
On the other hand, the second peak position, 
which is governed by $\beta_{+}=-2(J-H)$, shifts towards to $N_{d}=N$.    
The area of the
second peak, which is calculated  as in eq.(\ref{pall}), approaches $P_{d}$ 
and the width becomes narrow.   
$P(N,\rho_{d})$ increases monotonically to $P_{d}$ with $\rho_{d}$ and  
$P(N_{d},\rho_{d})$ near $N_{d}=N$ is upward convex.

 We would like to discuss the above effect on the tranche 
of securities synthesized from  the homogeneous assets pool with
parameters $P_{d},\rho_{d}$. 
For the purpose, it is useful to 
introduce the cumulative distribution function $D(i,\rho)$, which is
defined as
\begin{equation}
D(i,\rho_{d})=\sum_{N_{d}=i}^{N}P(N_{d},\rho_{d})  .
\end{equation}
From the definition $D(0,\rho_{d})=1$ and 
$D(1,\rho_{d})=1-P(0,\rho_{d})$ is the probability of the occurrence 
of default. 
We explain the 
relation between $D(i,\rho_{d})$
 and the evaluation of the tranche briefly.

The tranche for the interval $[i,j]$ implies that if the number of 
default $N_{d}$ is below $i \hspace*{0.2cm}(N_{d} <i)$, 
the tranche does not suffer from any
damage. However, if $N_{d}$ exceeds or becomes equal to $i 
\hspace*{0.2cm}(N_{d}\ge i)$,  it begins to lose its value.
The value of the tranche is  $\Delta=j-i+1$ in 
units of the number of assets (we assume that the 
values of all assets in the pool are equal.) 
and if defaults with $i \le N_{d} \le j$
occurs, it loses $(N_{d}-i+1)$ units. When $N_{d}$ exceeds $j
(N_{d}>j)$, the tranche lose its value completely.
The expected loss rate of the tranche $[i,j]$ is calculated as
\begin{equation}
E(i|j)=\frac{1}{\Delta}
\left(\sum_{k=i}^{j}
P(k,\rho_{d})(k-(i-1))
+\Delta \sum_{k=j+1}^{N}P(k,\rho_{d})\right).
\end{equation}
The first terms comes from the partial damage in the tranche
$(i \le N_{d} \le j)$ and the second term implies the contribution from
its complete loss of the tranche $(N_{d}\ge j+1)$.
$E(i|j)$ are directly related with the price of the tranche (premium),
which can be observed in the market.
For $j=i$, we denote $E(i|i)$ as $E(i)$ and call it as the expected loss
rate at the $i$-th tranche. 
It is related with the cumulative distribution $D(i,\rho)$ as
\begin{equation}
E(i)=E(i|i)=
\sum_{k=i}^{N}P(k,\rho_{d})=D(i,\rho_{d}).
\end{equation}
$E(i)$ is useful, because we can reconstruct $E(i|j)$ as a 
sum of $E(k)$ as
\begin{equation}
E(i|j)=\frac{1}{\Delta}
\sum_{k=i}^{j}E(k).   \label{E1}
\end{equation}
The proof of the relation is straightforward.
\begin{eqnarray}
\sum_{l=i}^{j}E(l)&=&\sum_{l=i}^{j}(\sum_{k=l}^{N}
P(k,\rho_{d}))=\sum_{l=i}^{j}(\sum_{k=l}^{j}
P(k,\rho_{d})+\sum_{k=j+1}^{N}
P(k,\rho_{d})) \nonumber \\
&=& \sum_{k=i}^{j}P(k,\rho_{d})(k-(i-1))
+\Delta \sum_{k=j+1}^{N}P(k)
=\Delta E(i|j)  \label{E2}
\end{eqnarray}
We note that, if we set $i=1$ and $j=N$ in equation \eref{E1}, we obtain 
\begin{equation}
E(1|N)=\frac{1}{N}\sum_{k=1}^{N} E(k)=P_{d}. \label{E3}
\end{equation}
Here we use the relation $E(1|N)=P_{d}$, which is 
intuitively clear and can be proved as in the equation \eref{E2}.
From the second equality in eq.\eref{E3} 
that the average of the expected loss rate at each tranche is $P_{d}$,
 tranches look like to ``share $P_{d}$ between them'' 
or ``toss $P_{d}$ to other tranches''.

\begin{figure}[htbp]
 \begin{center}
  \includegraphics[width=0.9\linewidth]{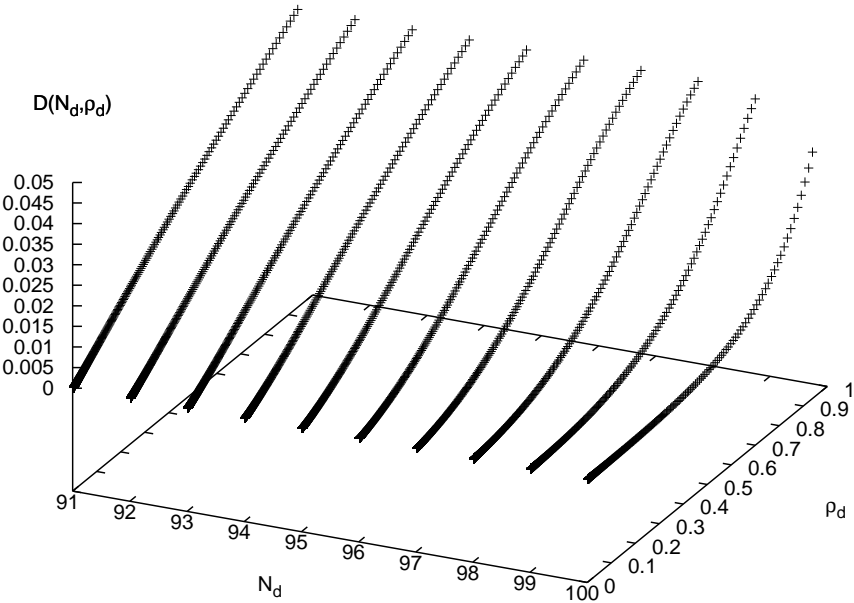}
  \caption{Plot of $D(i,\rho_{d})$ vs $\rho_{d}$. $ 91 \le i \le 100$.}
  \label{fig:91to100D}
 \end{center}
\end{figure}

\begin{figure}[htbp]
 \begin{center}
  \includegraphics[width=0.9\linewidth]{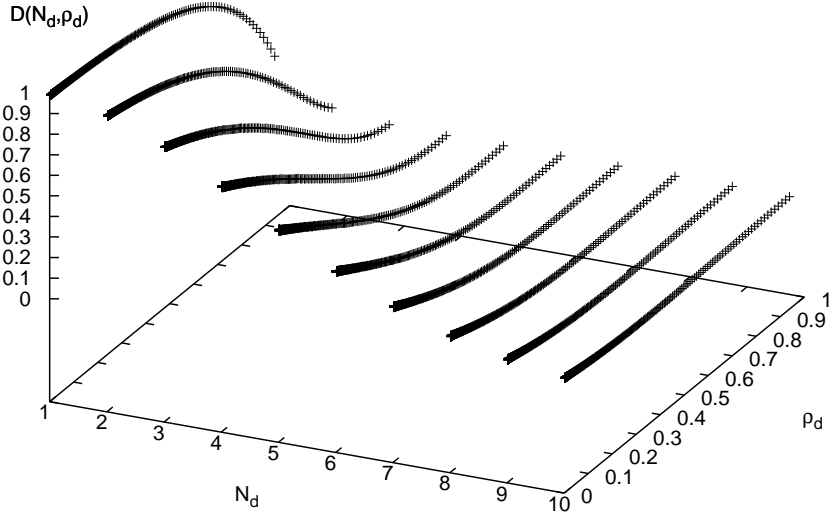}
  \caption{Plot of $D(i,\rho_{d})$ vs $\rho_{d}$. $D(i,\rho_{d})
=\sum_{k=i}^{N}P(k,\rho_{d})$ and $1\le i \le 10$.}
  \label{fig:1to10D}
 \end{center}
\end{figure}

\begin{figure}[htbp]
 \begin{center}
  \includegraphics[width=0.9\linewidth]{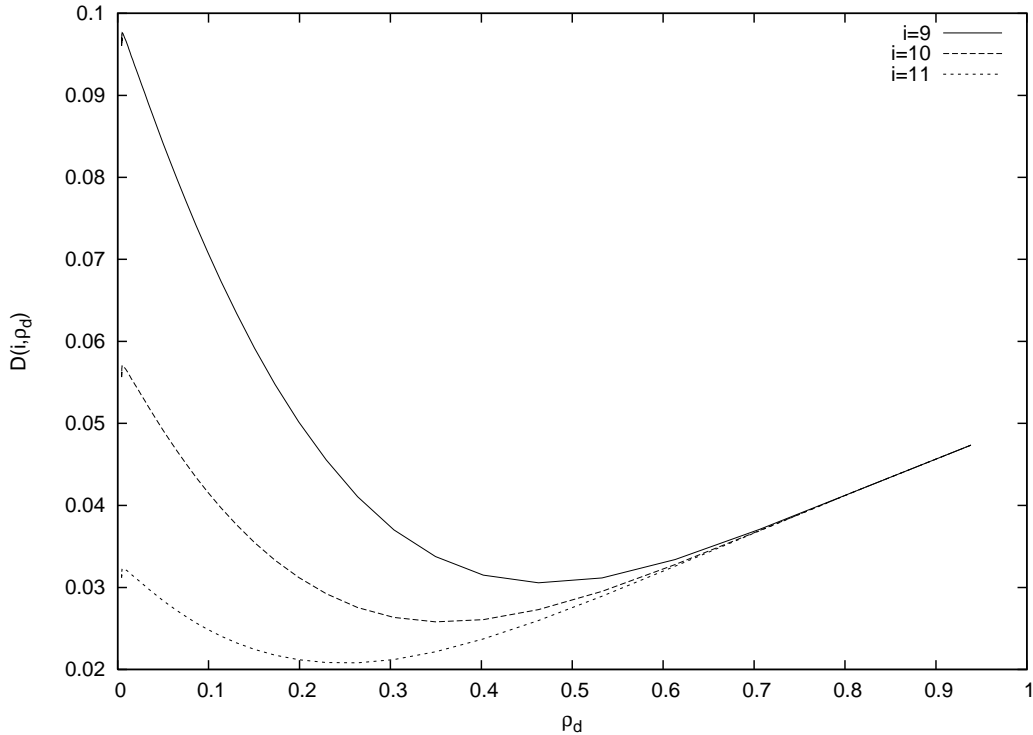}
  \caption{Plot of $D(i,\rho_{d})$ vs $\rho_{d}$. $ 9 \le i \le 11$.
  Solid line $i=9$, broken line $i=10$ and dotted line $i=11$.}
  \label{fig:9to11D}
 \end{center}
\end{figure}

Now we discuss the effect of the default correlation $\rho_{d}$ on 
$E(i)=D(i,\rho_{d})$. From the definition 
$D(i,\rho)=\sum_{k=i}^{N}P(k,\rho_{d})$, 
we can understand the
$\rho_{d}$ dependence easily from the previous discussions on 
$P(N_{d},\rho_{d})$. In \fref{fig:91to100D}, we show 
$D(i,\rho_{d})$ for $91 \le  i \le 100$. 
The area of the second peak
increases monotonically to $P_{d}$ as we increase $\rho_{d}$,
the cumulative distributions $D(i,\rho_{d})$ also grow up to $P_{d}$.
$P(N_{d},\rho_{d})$ is almost zero for $11 \le N_{d} \le 90$, $D(i,\rho_{d})$
for $11 \le i \le 90$
 behaves in the same way with these $D(i,\rho_{d})$ for $i=91$. 
If $i$ becomes
 small, we see the contribution from the first peak in
 $P(N_{d},\rho_{d})$. For small $i$, the damps of $P(N_{d},\rho_{d})$
 for $N_{d} \neq 0$
 with respect to $\rho_{d}$ dominates the contribution from the second
 peak. $P(0,\rho_{d})$ increases monotonically and
$D(1,\rho_{d})=1-P(0,\rho_{d})$ decreases with $\rho_{d}$. 
$D(i,\rho_{d})$ for $i\ge 2$ also decreases as in 
\fref{fig:1to10D}, which shows $D(i,\rho_{d})$ for $1\le i \le 10$.
These behaviors reflect the left shift and the width tinning 
of the first peak. For the intermediate value of $i$, 
the $\rho_{d}$ dependence of $D(i,\rho_{d})$ 
is not monotonous. In \fref{fig:9to11D}, we
 depict $D(i,\rho_{d})$ for $9 \le i \le 11$. 
Along with the shape change of the first peak 
with $\rho_{d}$, $D(i,\rho_{d})$
at first decrease. Then, the contribution from the second peak dominates
the decrease of the first peak contribution and 
$D(i,\rho_{d})$ begins to increase. 
$D(i,\rho_{d})$ is downward convex
 with respect to $\rho_{d}$ for the interval of $i$. 

We note that the 
ranges where $D(i,\rho_{d})$ is downward convex, 
$D(i,\rho_{d})$ decreases monotonically, 
or $D(i,\rho_{d})$ increases monotonically depends on the 
parameters $N,P_{d}$. The above discussions 
may not hold for other values of $N$ and $P_{d}$. 
In particular the range of the downward convex region,
 if we set $P_{d}=0.01$, we observe that it shift to the left.
The positions of the boundaries between the 
regions are important  from the
viewpoints of risk management and rating of the securities,
 we should note this point.

From these observations, we summarize the $\rho_{d}$ dependence of $E(i)
=D(i,\rho_{d})$.

\begin{itemize}

\item Senior tranche, the range $[i,j]$
with  $i,j$ are set to be large, 
suffers from the default 
correlation seriously. $E(i)$ for the range increase linearly 
with $\rho_{d}$. It is approximately given by
\begin{equation}
E(i) =\rho_{d}\times P_{d}
\end{equation}
If $\rho_{d}$ change from $0$ to $1$, $E(i)$ change from $0$ to $P{d}$
 and the evaluations of the securities decrease almost linearly.  

\item Equity or subordinate tranche, the range $[i,j]$
is $[1,j]$ and $j$ is small. In the range
$1 \le k \le j$, $D(k,\rho_{d})$ damps monotonically with $\rho_{d}$
and $E(1|j)$ also damps. The increase in $\rho_{d}$ causes the increase
of the evaluation of the tranche.  

\item Mezzanine tranche, whose range is between the equity and the senior
      tranche. 
In the range, the behaviors of $D(k,\rho_{d})$ depends on $P_{d}$ and
      the system size $N$. In the above mentioned $N=100,P_{d}=0.05$ case, 
$D(k,\rho_{d})$ has downward convex shape for some intermediate values
 of $k$.
 
\end{itemize}

\section{Concluding Remarks and Future Problems}
\label{conclusion}

In this paper, we have studied the long-range Ising  model as 
a model for a pool of $N$ homogeneous assets with default probability $P_{d}$
and default correlation $\rho_{d}$. We have studied in the $(J,H)$
plane, the behavior of $P_{d}$ and $\rho_{d}$. There are two phases in
the $(J,H)$ plane. In the One Peak Phase, the probabilities $P(N_{d})$
have a single peak at $N_{d} \simeq N  P_{d}$. The correlation
is almost zero in the phase. In the Two Peak Phase, there are two peaks
in $P(N_{d})$ and $\rho_{d}$ can take large value. The first peak
is closer to origin than $N P_{d}$ and its area is larger than
$1-P_{d}$. The second peak is at about $N$ and its area is less than $P_{d}$.
The parameters $J,H$
should be chosen in the Two peak phase, if the model intends to describe
the portfolio with some default correlation between the 
assets. We have developed the perturbative 
method and expressed $P(N_{d})$  as a superposition of 
two binomial distributions with the above two peaks at zero-th order.
We have obtained the closed form expression for $P_{d},\rho_{d}$ and the
weights for the second peak $P_{all}$, which means the probability 
that almost all assets are defaulted. These expressions are in good
agreement with numerically calculated values and give an efficient method
 for the actual application of the long-range Ising model. Otherwise,
 for $P_{d}$ and $\rho_{d}$, it is difficult to know the parameters $J$
 and $H$ and the long range 
Ising model is hard-to-use as a model for homogeneous credit risk
 portfolio. 
 
Furthermore, we have studied the $\rho_{d}$ dependence of 
$P(N_{d},\rho_{d})$ and
the cumulative distribution $D(i,\rho_{d})$. $P(N_{d},0)$ is binomial
distribution $B(N,P_{d})$ and it has a peak at $N P_{d}$.
As we increase $\rho_{d}$ from
0 to 1, the profile of $P(N_{d},\rho_{d})$ changes from One peak shape
to Two Peak shape. The first peak
shifts to the left and its shape 
becomes higher and narrower. The second peak's area increases and it
shifts to the right with the decrease of its width.  
At $\rho_{d}=1$, $P(N_{d},1)$ 
has two thin peaks at $N_{d}=0$ and $N_{d}=N$ and  the probabilities
are $P(0,1)=1-P_{d}$ and $P(N,1)=P_{d}$. Other probabilities 
are zero. The cumulative distribution functions $D(i,\rho_{d})$
 correspond to  the average loss rates of the $i$-th tranche.
About the senior tranche, the range of the tranche $[i,j]$ 
is large.
As $\rho_{d}$ increases, $D(i,\rho_{d})$ 
increase almost linearly with $\rho_{d}$
like  $D(i,\rho_{d}) \simeq \rho_{d}\times P_{d}$. 
The average loss rate of the senior tranche $[i,j]$ is given as a sum of 
$D(k,\rho_{d})$ in the range $[i,j]$, the expectation
value of the loss rate of the senior tranche also increases as
$\rho_{d}\times P_{d}$. The  price of the tranche is based on the 
average loss rate, the value of the senior tranche
decreases with $\rho_{d}$. 
 The range of the equity, the subordinated
tranche, is near the origin and the $D(i,\rho_{d})$s decrease
monotonically.
The average loss rate of the equity decreases with $\rho_{d}$ and
the price of the equity increase with $\rho_{d}$.
The mezzanine tranche is between the equity and the senior tranche.
The profile of $D(i,\rho_{d})$ in the range depends on the model parameters
$\rho_{d},P_{d}$ and $N$. In the text example, $D(i,\rho_{d})$ has a
downward convex shape in some region. If the mezzanine range $[i,j]$ 
is chosen to lie in the region, the average loss rate also behaves similarly.  
However, other probabilistic model for a pool of assets, e.g. the copula
model \cite{Li}, 
suggest
upward convex shape for the average loss of the mezzanine tranche.
The discrepancy comes from the difference of the shapes of  
$P(N_{d},\rho_{d})$. The more complete comparison between the
probabilistic models for a pool of correlated assets should be 
done.

As concluding remarks, we comment on the usage of the long-range Ising
model and related future problems. As a statistical 
model for an ensemble of many assets, the long-range Ising model is
an attractive one from the viewpoint of physicists. Its phase diagram
and phase transitions are throughly studied and its analytic calculation
method, like Hubbard-Stratanovich transformation, guides us how to 
make theoretically tractable models. On the other hand, from the 
viewpoint of financial engineers, the long-range Ising model is not
so convenient. One reason is that the model parameters $J,H$ are not
directly related with the observed data $P_{d}$ and $\rho_{d}$ (or
$\rho_{a}$).  Other statistical models incorporate these parameters
 as a model parameters. For example, the Moody's correlated binomial
model gives $P(N_{d})$ as a function of $P_{d}$ and $\rho_{d}$ 
explicitly.
When one uses Ising model, it is necessary
to know the parameters $J,H$ which correspond to $P_{d},\rho_{d}$. 
The definition of $P_{d}$ and $\rho_{d}$
include the moments $<N_{-}^{l}>$ or $<S_{i}>$ 
and $<S_{i}S_{j}>$, it is necessary to take the trace $\Tr$.
The long-range
Ising model has the advantage that the trace $\Tr$ is reduced to the summation
over the total magnetization $M=\sum_{j=1}^{N}S_{j}$ and the calculation
is not so heavy task. Even so, this one step spoils the usefulness of
the model. We have obtained a closed form expressions for $P_{d}$ and
$\rho_{d}$ and try to circumvent the step. The computational time 
to obtain $J,H$ for given $P_{d},\rho_{d}$ is reduced much and 
the failing of the model are partially overcome.   

In order to apply the long-range Ising model to 
the evaluation of the tranche $[i.j]$ in more realistic situation, 
the assumption of homogeneity of the assets pool
should be weakened. One step toward the direction is to introduce 
multi sectors and assume the homogeneity only in each sector.
We label each sector by $I=1,2,\cdots K$ and $I$-th sector contains
$N^{I}$ assets. In the $I$-th sector, the default rate is $P_{d}^{I}$
and the default correlation is $\rho_{d}^{I}$. 
Between different sectors, say between $I$-th and $J$-th sector,
the default correlation is $\rho_{d}^{IJ}$.
We use Ising Spin variables $S^{I}_{j}$ to represent the states
of the $j$-th asset in the $I$-th sector, the generalized long-range
Ising model Hamiltonian for the probabilities $P(S^{I}_{j})$ is 
\begin{equation}  
-{\cal H}=
\sum_{I=1}^{K}
\frac{J^{I}}{2 N^{I}}M_{I}^{2}+H^{I}M_{I}
+\sum_{1=I<J=K}\frac{J^{IJ}}{\sqrt{N^{I}N^{J}}}
M_{I}M_{J}
\end{equation}
As in the homogeneous model, the Hamiltonian depends on $S^{I}_{j}$
only through the magnetization of the $I$-th sector $M_{I}=
\sum_{j=1}^{N^{I}}S^{I}_{j}$. When we set $\frac{J^{I}}{N^{I}}
=\frac{J^{IJ}}{\sqrt{N^{I}N^{J}}}=\frac{J}N$ 
and $H^{I}=H$, the model reduces to the homogeneous model with 
$N=\sum_{I=1}^{K}N^{I},J,H$. The problem is to get the relation between
$\rho_{d}^{I},\rho_{d}^{IJ},P_{d}^{I}$ and $J^{I},J^{IJ},H^{I},N^{I}$.
In order to accomplish the task, the phase diagram in 
$J^{IJ},J^{I},H^{I}$ and the profile $P(N^{I}_{d})$ 
should be cleared and it is left for future analysis.
Furthermore, for more complex situation where $i$-th asset has 
default probability $P_{d}^{i}$ and the default correlation between
$i$-th and $j$-th asset is $\rho_{d}^{ij}$, the model Hamiltonian
becomes that of the random Ising spin systems. The exchange interaction
$J_{ij}$ and the external field $H_{i}$ should be connected to 
$P_{d}^{i}$ and $\rho_{d}^{ij}$, which is also left for future problem.
Other step is to discard the Ising model and adopt other
probabilistic models. One possibility is the Moody's 
correlated binomial model, which uses two state variables $x=0,1$
for the state of an asset
and incorporates
$P_{d}$ and $\rho_{d}$ directly in the model parameters. Its
generalization to the multi-sector case and more complex situations 
is an interesting problem. 
Other possibility is to introduce simplified version of  
the long-range Ising model. 
We  use two state variable $x_{i}$ for the state of the $i$-th asset.
The number of defaults $N_{d}$ is expressed as $N_{d}=
\sum_{j=1}^{N}x_{j}$. 
The probabilities $P(x_{1},x_{2},\cdots,x_{N})$ is given as
\begin{equation}
P(x_{1},x_{2},\cdots,x_{N})=
(1-\alpha) p^{N_{d}}(1-p)^{N-N_{d}}+\alpha \delta_{N,N_{d}}.
\end{equation}
Instead  of the superposition of two binomial distributions,
we use $\alpha \delta_{N,N_{d}}$ for the second peak. The first peak 
is $B(N,p)$ and the parameters $\alpha, p$ are related with  the default
probability $P_{d}$ as $P_{d}=\alpha +(1-\alpha)p$.
This probabilities $P(x_{1},x_{2},\cdots,x_{N})$ 
is more tractable than the original probability distribution \eref{Ising} 
and the
generalizations to more complex situations may be carried out easily.

\ack{
This work has received financial support from Kitasato University,
project SCI:2005-1706.}

\end{document}